\documentclass[a4paper,12pt]{article}

\usepackage{epsfig}
\usepackage{a4}
\usepackage{latexsym, amssymb} 
\usepackage{amsmath}
\newcommand{\lapprox}{%
\mathrel{%
\setbox0=\hbox{$<$}
\raise0.6ex\copy0\kern-\wd0
\lower0.65ex\hbox{$\sim$}
}}
\newcommand{\gapprox}{%
\mathrel{%
\setbox0=\hbox{$>$}
\raise0.6ex\copy0\kern-\wd0
\lower0.65ex\hbox{$\sim$}
}}

\newcommand{\ba}{\begin{array}}
\newcommand{\ea}{\end{array}}
\newcommand{\bd}{\begin{displaymath}}
\newcommand{\ed}{\end{displaymath}}
\newcommand{\be}{\begin{equation}}
\newcommand{\ee}{\end{equation}}
\newcommand{\bea}{\begin{eqnarray}}
\newcommand{\eea}{\end{eqnarray}}

\def\fb{\, {\rm fb}}

\def\q2 {q^2}

\def\bt{\begin{table}}
\def\et{\end{table}}

\catcode`@=11 
\def \gsim{\mathrel{\mathpalette\@versim>}}
\def \lsim{\mathrel{\mathpalette\@versim<}}
\def \@versim#1#2{\lower0.4ex\vbox{\baselineskip\z@skip\lineskip\z@skip
     \lineskiplimit\z@\ialign{$\m@th#1\hfil##\hfil$%
     \crcr#2\crcr\sim\crcr}}}
\catcode`@=12 

\begin{document}

\begin{flushright}
{\small 
RECAPP-HRI-2012-008}
\end{flushright}

\begin{center}

{\large\bf New Higgs interactions and recent data from
the LHC and the Tevatron}\\[15mm]
Shankha Banerjee\footnote{E-mail: shankhaban@hri.res.in},
Satyanarayan Mukhopadhyay \footnote{E-mail: satya@hri.res.in}
and Biswarup Mukhopadhyaya\footnote{E-mail: biswarup@hri.res.in}\\
{\em Regional Centre for Accelerator-based Particle Physics \\
     Harish-Chandra Research Institute\\
Chhatnag Road, Jhusi, Allahabad - 211 019, India}
\\[20mm] 
\end{center}

\begin{abstract} 
We perform a multi-parameter global analysis of all data available till date
from the ATLAS, CMS and Tevatron experiments, on the signals of a Higgs boson,
to investigate how much scope exists for departure from the standard model
prediction. We adopt a very general and model-independent scenario, where
separate deviations from standard model values are possible for couplings of
the observed scalar with up-and down-type fermions, W-and Z-boson pairs, as
well as gluon and photon pair effective interactions. An arbitrary phase in
the coupling with the top-pair, and the provision for an invisible decay width
for the scalar are also introduced. After performing a global fit with seven
parameters, we find that their values at 95\% confidence level can be
considerably different from standard model expectations. Moreover, rather
striking implications of the phase in top-quark coupling are noticed.  We also
note that the invisible branching ratio can be sizeable, especially when the
couplings of the Higgs to W-and Z-pairs are allowed to be different.
\end{abstract}

\vskip 1 true cm

\newpage
\setcounter{footnote}{0}

\def\baselinestretch{1.5}
\section{Introduction}
The recent announcements on Higgs boson search from the 2012 run of the Large
Hadron Collider (LHC)~\cite{CMSICHEP2012,ATLASICHEP2012}, in conjunction with
earlier results, have brought a refreshing gale into the stultified arena of
particle physics, starved of breakthrough for several decades.  Physicists
tend to concur that both the CMS and ATLAS collaborations have found a
spin-zero bosonic resonance of mass 125 - 126 GeV.  However, the issue of
whether it is exactly the Higgs boson predicted by the standard electroweak
model (SM)~\cite{GG2,Higgs-original} is still somewhat open. This question can
be settled by the close examination of accumulating data over a longer period,
thus revealing not only the complete picture of the electroweak symmetry
breaking sector but also any secret message of new physics buried within the
available results.

Any departure from SM predictions in the observed scalar is ultimately
reflected in its interactions with pairs of fermions or weak gauge bosons, and
also the loop-induced effective couplings to photon and gluon pairs.  However,
these interactions are rather intricately twined in the calculation of rates
for various final states into which the Higgs can decay.  The trace of
non-standard physics there can be looked for in two ways--- either in the
context of specific theories where the different new couplings are all
dictated by the model parameters, or in a phenomenological, model-independent
analysis of as general a nature as possible. We attempt to add the present
investigation in the second direction, to the already growing volume of extant
studies in the light of the data piling
up~\cite{Contino,fits-past,fits-recent}.

We start by assuming that the $SU(2)_L\times U(1)_Y$ gauge invariance holds at
the energy scale under scrutiny.  Under such circumstances, the aforementioned
interactions of the Higgs boson can be different from the SM expectations due
to (a) the observed scalar having admixtures with other states, or (b) the
presence of additional, gauge invariant effective operators which contribute
to the same couplings. In either case, a renormalisation of the SM couplings
may take place, and interactions with new Lorentz structure may also
appear. Only the former possibility is addressed in the present work.

The modification of the SM couplings of the Higgs is subject to various
experimental constraints, the severest of them often coming from electroweak
precision observables. Our purpose is to investigate, within such constraints,
how much allowance for departure of various interaction strengths from the SM
values can be made by the currently observed rates in different final states,
as found from a global fit of these strengths based on chi-square
minimisation. The available best fits for the rates in different channels,
scaled by the corresponding SM expectations, are our inputs, and we use them
to obtain the widest allowed ranges for various modified couplings as well as
an invisible decay width, with the hope that this will guide us to the
direction where new physics may lie.

We start by assuming that the Higgs (or, to be more precise, the observed
boson, to which we refer here as the Higgs) can differ from the SM
expectations in all respects--- couplings with $T_3 = + 1/2$ and $T_3 = -1/2$
fermions (the departures being different in the two cases), couplings with
both W-and Z-boson pairs (again with potentially different deviations), and
also the effective couplings to photon and gluon pairs, where additional
effects over and above the modified fermion and boson couplings are included
as possibilities.  While the modifications mentioned above can be described in
terms of a given framework (such as a chiral Lagrangian), we deliberately take
them as completely free and independent parameters, to make our study free of
any theoretical bias. Besides, it is also assumed at the beginning that the
Higgs can have a finite invisible decay width.  We use the CMS as well as
ATLAS results in various channels for both the 2011 (7 TeV) and 2012 (8 TeV)
runs~\cite{LHC-papers}. Results from the Tevatron, wherever available, are
also used as data points in obtaining the least-square fits~\cite{Tevatron}.

As has been already mentioned, a number of similar investigations have
appeared in the literature~\cite{fits-recent}. Though they are all
instructive, the present study may be of particular use in the following
respects:

\begin{itemize}
\item We start by taking {\em all} of the couplings, tree-level as well as
  loop-induced, to be unrelated and free parameters. The couplings to up-and
  down-type quarks, as also those to W-and Z-pairs, are allowed to be
  different at the same time. Most earlier studies (except, for example, 
  Refs.~\cite{Contino} and~\cite{DysZphilia}) have not allowed
  uncorrelated variation of $HWW$ and $HZZ$ interactions, albeit in a
  gauge-invariant fashion; also, some of them set the interactions with up-and
  down-type quarks to identical value.

\item We investigate the effect of an arbitrary phase in the fermion pair
  couplings, relative to the W-pair couplings, which is also varied as a free
  parameter in obtaining the global fits.

\item We do not restrict ourselves by assuming the anomalous interactions to
  arise from some underlying scenario like a chiral Lagrangian, which implies
  concomitant variation of different couplings.
\item We take into account the possibility that the effective $Hgg$ and
  $H\gamma\gamma$ couplings are modified due to effects other than
  non-standard interactions of the Higgs with W-and fermion
  pairs. Furthermore, the possibility of loop-induced Higgs decays being
  modified differently due to both coloured and colourless fermions running in the
  loops is taken into consideration. We keep the modifications due to these
  two different effects separate and uncorrelated.

\item An invisible decay width for the Higgs is allowed as a free parameter.
\item We parametrize all new physics effects at the coupling level (excepting
  for the invisible width, where we take the width as a free parameter, for
  the sake of model--independence). This is in contrast to cases where
  branching ratios of the Higgs into some channels are taken as free
  parameters, which results in various new physics effects getting entangled,
  since branching ratios also involve the total decay width of the Higgs,
  where all (modified) interactions contribute.

\item In obtaining the $2 \sigma$ ranges of allowed values for various
  couplings, we have fixed the remaining parameters not at their SM values but
  at their global best fits corresponding to the minimum of the $\chi^2$
  function. This makes our analysis completely unbiased, from the angle of
  physics beyond the SM.

\item Even in channels where the gluon fusion channel for Higgs production
  dominates, the vector boson fusion (VBF) and associated production channels
  contaminate the rates to varying degrees. Also, the VBF and associated
  channels have both $HWW$-and $HZZ$-induced contributions. These warrant a
  careful treatment if one is allowing uncorrelated deviations from SM in the
  $HWW$ and $HZZ$ interactions.  Difficulties on this front are compounded by
  different efficiency of cuts in different channels. Such issues have been
  taken into account in our analysis.
\item In cases where only the 2011 data are used, where the best
fit values for various rates come with asymmetric error-bars, the
asymmetry is retained in the analysis that follows. 
\end{itemize}

In section 2, we outline our parametrization of the various new physics
effects mentioned above, and discuss the motivations of their origin. The
details of input data sets from the LHC and the Tevatron used in our global
numerical analysis, and the methodology adopted have been described in section
3. We discuss our results of the best-fit values obtained and the allowed
confidence intervals for the various parameters in section 4.  We summarise
and conclude in section 5.

\section{New physics effects: parametrization}

In this section, we describe the parameters used to encapsulate any likely new
physics effect hidden within the data on the Higgs. We also outline the
motivations for choosing these parameters, and indicate the ranges over which
we let them vary to obtain their favoured locations in the light of
observations till date.

\vspace{0.4cm}
{\underline {\bf Fermion couplings}}

Classifying all $T_3 = +1/2$ fermions as u, and 
all $T_3 = -1/2$ fermions as d,  we assume

\begin{align}
{\cal L}^{eff}_{H{\bar u} u} &= e^{i\delta}\alpha_u\frac{m_u}{v}H\bar{u}u \nonumber \\
{\cal L}^{eff}_{H{\bar d} d} &= \alpha_d\frac{m_d}{v}H\bar{d}d
\end{align}

\noindent
where $H$ denotes the scalar with a Higgs-like appearance. This
parametrization implicitly assumes the couplings of $H$ to all fermions to be
proportional to their masses. The assumed difference in interaction strengths
to up-and down-type fermions include the possibility of $H$ being part of a
scenario containing more than one doublets, including the supersymmetric case.
In general, a relative phase between the couplings is assumed~\footnote{A phase in the $H t \bar{t}$ effective coupling can arise due to imaginary
(absorptive) parts coming from loop diagrams for the transition where some of the
intermediate SM states in the loop graphs, being lighter than the
Higgs boson, can go on-shell. For example, a heavy $W^\prime$ like gauge
boson having $W^\prime t b$ type couplings can give rise to additional
contributions to the $H  t \bar{t}$ effective coupling, via a triangle loop
involving two b-quarks, where the b-quarks can go on-shell inside the
loop. This would then give rise to an imaginary part in the effective interaction.
}.  This phase
allows various degrees of interference among fermion loops, and between the
fermion and W-boson loops, for example, in the decay $H \longrightarrow
\gamma\gamma$. As will be seen in the next section, we perform a scan over the
phase $\delta$ like we do over $\alpha_u$ and $\alpha_d$. Even in cases where
we neglect a non-trivial phase, we allow $\alpha_u$ and $\alpha_d$ to be both
positive and negative, in order to account for constructive as well as
destructive interferences.  Moreover, the phase $\delta$ enters seriously into
Higgs phenomenology only via the top quark couplings. Therefore, nothing is
affected by dropping it for the first two families, in case it is subject to
any constraints from flavour physics.

There are essentially no limits on $|\alpha_u|$ and $|\alpha_d|$ from earlier
data, except those from perturbativity of the Yukawa couplings. Keeping this
in mind, we take the maximum value of $|\alpha_u|$ to be 2 (keeping the top
quark Yukawa coupling in mind), while $|\alpha_d|$ is {\it prima facie}
allowed to lie all the way upto 40.  However, even before the full analysis is
done, we find that the rate for $H\longrightarrow \gamma\gamma$ for large
$|\alpha_d|$ gets suppressed well below limits admissible by current
observations, as the $b{\bar b}$ mode then dominates overwhelmingly.  With
this in view, we also limit the maximum value of $|\alpha_d|$ to 2 in our
final analysis. Since this renders the contribution of all $T_3 = -1/2$ SM
fermions to $\Gamma (H\longrightarrow \gamma\gamma)$ negligibly small, we also
do not miss anything by dropping the phase in $\alpha_d$. Note that the same 
consideration prevents us from taking seriously regions of the parameter space, 
where the b-quark loop contributes substantially to the gluon fusion channel
of Higgs production. Thus we do not expect the phase in $H b \bar{b}$ coupling,
too, to affect Higgs production cross-sections.

\vspace{0.4cm}
{\underline {\bf Gauge boson pair couplings}}

We parametrize the interactions of the
observed scalar to a pair of weak gauge bosons as

\begin{align}
{\cal L}_{HWW}^{eff} &= \beta_W\frac{2m_W^2}{v}HW_{\mu}^+W^{\mu-} \nonumber\\
{\cal L}_{HZZ}^{eff} &= \beta_Z\frac{m_Z^2}{v}HZ_{\mu}Z^{\mu}
\end{align}

\noindent
where the Lorentz structure of the interaction has been tentatively taken to
be the same as in the standard model.  It should be emphasized that this
parametrization, especially allowing $\beta_W \ne \beta_Z$, allows one to
address the apparent suppression of the $WW$ channel as opposed to the $ZZ$
channel, as evinced in the 2011 data.  Such anomalous interactions can arise
if, again, the Higgs has admixtures of other doublets or scalars in some other
representations of SU(2), and also via loop effects in specific models.

Clearly, one faces precision electroweak constraints here, in particular, for
$\beta_W \ne \beta_Z$, which entails a breakdown of custodial SU(2), and is
thus restricted by the T-parameter. Such anomalous couplings can arise, for
example, from gauge invariant effective operators, an example
being~\cite{Hagiwara}

\be
{\cal O_\phi} = {\frac {f_\phi}{\Lambda^2}} (D_\mu \Phi)^\dagger  \Phi \Phi^\dagger (D_\mu \Phi)
\ee

\noindent
where $\Phi$ is a set of SU(2) doublet scalars, out of which $H$ is the
lightest mass eigenstate, and $\Lambda$ is the scale below which the effective
operator is defined. This operator in itself gives rise to unequal $\beta_W$
and $\beta_Z$.  However, taking this operator alone, precision constraints
yield the limits~\cite{Hagiwara,Garcia}

\bea
0.991 \lsim \beta_W \lsim 1.001 \label{bound-1}\\
0.997 \lsim \beta_Z \lsim 1.028 
\label{bound-2}
\eea

It is hardly expected to see any appreciable effects of such variation on
observable rates, and we do not include the bounds given by Eqns.~\ref{bound-1},~\ref{bound-2} 
in our global fits.  However, one can have less constrained couplings if one
includes other effective operators which, however, give rise to additional
$HWW$ and $HZZ$ interactions involving derivatives. We do not rule out such
possibilities, but neglect the effects of derivative couplings for the time
being, and vary $\beta_W$ and $\beta_Z$ between $0$ and $2$, in a purely
phenomenological way. An analysis including the derivative couplings will be
presented in a subsequent study. We also consider the case where $\beta_W =
\beta_Z \equiv \beta$, thereby restoring tree-level custodial invariance. In
this case, for a Higgs mass of $125$ GeV, electroweak precision constraints
restrict $\beta$ in the range~\cite{Contino}

\be 
0.84 \leq \beta^2 \leq 1.4 
\ee

\vspace{0.4cm}
{\underline {\bf Effective gluon-gluon and photon-photon couplings}}

The gluon fusion channel is the dominant production mode for a Higgs of mass
around 125 GeV, and is overwhelmingly driven by the top quark loop.
Therefore, a departure of $\alpha_u$ from unity will be reflected in the
production cross-section (though the phase $\delta$ will be
ineffectual). Similarly, the two-photon amplitude, which is responsible for
perhaps the most discussed final state nowadays, has contributions from both
fermion-and $W$-induced loops. Thus the parameter $\beta_W$ also dictates the
rate for the two-photon final state.

This is, however, not the entire story. Both of the aforementioned
loop-induced processes can have modified contributions, beyond the coverage of
the $\alpha$-and $\beta$-parameters, if additional states contribute in the
loops.  The most obvious example is the contribution of Kaluza-Klein towers in
theories with extra compact dimensions, where fermions and/or gauge bosons
propagate in the bulk. Due to such (and perhaps other) possibilities, it is
necessary in a general study to include an additional parameter to properly
quantify new physics effects in the gluon fusion channel. For the two-photon
amplitude, this parameter can well be different, since new physics can be
quite different in the coloured and non-coloured sectors.

With this in view, we parametrize the gluon-gluon-Higgs and
Higgs-photon-photon amplitudes as follows:

\begin{align}
{\cal L}_{gg}^{eff} &= -x_g f(\alpha_u)\frac{\alpha_s}{12\pi v}HG_{\mu \nu}^aG^{a
  \mu \nu} \nonumber\\ 
{\cal L}_{\gamma \gamma}^{eff} &= -x_{\gamma}g(\alpha_u,
\alpha_d, \beta_W, \delta)\frac{\alpha_{em}}{8\pi v}HF_{\mu \nu}F^{\mu \nu}
\end{align}

\noindent
where $x_g$ and $x_\gamma$ encapsulate the overall modification due to new
intermediate states in the two cases~\footnote{This is just one way of parametrizing 
the effects of new states. It could be parametrized alternatively adding terms of 
the form $x_g HG_{\mu \nu}^aG^{a \mu \nu}$ or $x_\gamma HF_{\mu \nu}F^{\mu \nu}$ to
the SM Lagrangian. It is straightforward to translate the limits obtained using one 
convention into those using the other.}.
The functions $f(\alpha_u)$ and
$g(\alpha_u, \alpha_d, \beta_W, \delta)$ encapsulate the modifications of
these couplings due to fermion and W-boson loops. We shall discuss their
detailed forms in the next section when we discuss the departures of the Higgs
production and decay widths from their SM values in detail. Since there is no
restriction till now on $x_g$ and $x_\gamma$ , we let each of them vary from
0.2 to 3.0. The lower and upper bounds have been set keeping in mind that we
do not have too much room for varying them beyond a range, and still being
consistent with the data.

\vspace{0.4cm}
{\underline {\bf Invisible width}}

Since the earlier analyses of the 2011 data have led to different conclusions
about a possible invisible width of the Higgs~\cite{fits-past}, we keep this
possibility alive in our analysis.  Such an invisible width can occur if, for
example, the Higgs serves as a portal to a `dark matter' sector.  The exact
expression of the width in terms of the coupling to the dark sector will
require one to know the nature of the invisible particle(s), for example,
whether they are scalars or spin-1/2 objects. In order to be
model-independent, we take as a free parameter the {\em invisible width}
$\Gamma_{inv}$, which is independent of the nature of the invisible state, and
is also not entangled with other new physics effects.

Since there is very little guideline on the range over which $\Gamma_{inv}$
should be varied in order to obtain the value corresponding to the minimum of
chi-squared, we start from $\epsilon$, the invisible branching ratio, and let
it vary between 0 and 1. In each case, the invisible width is expressed as

\be
\Gamma_{inv} = \frac{\epsilon}{1 -\epsilon} \sum{\Gamma_{vis}}
\ee

\noindent
where $\sum\Gamma_{vis}$ is the total decay width into 
all visible channels.

\section{Methodology of analysis}

\subsection{Input data}

Table 1 contains the details of all the data points used in our analysis. This
includes the combination of 7 TeV (with $5.1 \fb^{-1}$ data) and 8 TeV (with
$5.3 \fb^{-1}$ data) results from CMS~\cite{CMSICHEP2012,LHC-papers} in all
channels, namely, $\gamma\gamma$(inclusive), $ZZ^* \rightarrow 4\ell, WW^*
\rightarrow \ell \ell \nu \nu, \gamma\gamma jj, \tau^+\tau^-$ and
$b\bar{b}$. For the ATLAS experiment~\cite{ATLASICHEP2012,LHC-papers}, the
$\gamma\gamma$ and $ZZ^{*}$ results are available as similar $7+8$ TeV
combinations, whereas the 2011 results alone have been used for the remaining
channels. The integrated luminosities for ATLAS are $4.9 \fb^{-1}$ and $5.9
\fb^{-1}$ for the $7$ and $8$ TeV runs respectively. The $\gamma\gamma jj$
results from ATLAS are not yet available.  Furthermore, we have used the
Tevatron (combined CDF and D0) results for $WW^*, \gamma\gamma$ and
$b\bar{b}$, for an integrated luminosity of $10
\fb^{-1}$~\cite{Tevatron}. Thus we have fourteen input data points altogether
for our global analysis. All the SM production cross-sections and decay widths
for the Higgs have been taken from the results reported by the LHC Higgs Cross
Section Working Group~\cite{LHCHWG}.

\begin{table}[t]
\centering
\begin{tabular}{|l|c|c|c|c|}
\hline
Experiment & Channel & $\hat{\mu}$  & $M_H$ (GeV) & Energy (TeV)\\
\hline
\hline
 & $H \rightarrow W^{+}W^{-}$ & $0.32_{-0.32}^{+1.13}$  & & \\
Tevatron & $H \rightarrow b \bar{b}$ & $1.97_{-0.68}^{+0.74}$  & 125 & 1.96\\
 & $H \rightarrow \gamma \gamma$ & $3.62_{-2.54}^{+2.96}$  & & \\
\hline
 & $H \rightarrow \gamma \gamma$ & $1.9_{-0.50}^{+0.50}$  & 126.5 & $7 + 8$\\
 & $H \rightarrow Z Z^{*} \rightarrow 4l$ & $1.3_{-0.60}^{+0.60}$  & 126.5 & $7 + 8$\\
ATLAS & $H \rightarrow W W^{*} \rightarrow l^{+} l^{-} \nu \bar{\nu}$ & $0.52_{-0.60}^{+0.57}$ & 126 & 7\\
 & $H \rightarrow \tau \bar{\tau} $ & $0.16_{-1.84}^{+1.72}$  & 126 & 7\\
 & $H \rightarrow b \bar{b}$ & $0.48_{-2.12}^{+2.17}$  & 126 & 7\\
\hline
 & $H \rightarrow \gamma \gamma $ & $1.56_{-0.43}^{+0.43}$ &  & \\
 & $H \rightarrow Z Z^{*} \rightarrow 4l$ & $0.7_{-0.40}^{+0.40}$ & & \\
 & $H \rightarrow W W^{*} \rightarrow l^{+} l^{-} \nu \bar{\nu}$ & $0.62_{-0.45}^{+0.43}$ & & \\
CMS & $H \rightarrow \tau \bar{\tau} $ & $-0.14_{-0.73}^{+0.76}$ & 125.3 & $7 + 8$\\
 & $H \rightarrow b \bar{b} $ & $0.15_{-0.66}^{+0.73}$ & &\\
 & $H \rightarrow \gamma \gamma j j$ & $1.58_{-1.06}^{+1.06}$ & &\\ 
\hline
\end{tabular}
\caption{Input data set used in our analysis, with the values of $\hat{\mu_i}$
  in various channels and their $1\sigma$ uncertainties as reported by the
  ATLAS, CMS and Tevatron collaborations.}
\label{tab:tab1}
\end{table}

In calculating the modifications of various branching ratios of the Higgs, we
have used $m_t =$ 173.5 GeV, $m_b =$ 4.7 GeV, $m_{\tau} =$ 1.777 GeV and $m_W
=$ 80.4 GeV~\cite{PDG}. The best fit value for the Higgs mass ($m_H$) reported
by CMS is 125.3 $\pm$ 0.6 GeV~\cite{CMSICHEP2012}, whereas it is 126.5
GeV~\cite{ATLASICHEP2012} for ATLAS. The average of these two values, namely,
$125.9 $ GeV has been used by us, in line with, for example,
Ref.~\cite{Djouadi-recent}. For the Tevatron analysis, $m_H =$ 125 GeV has
been used, following combined CDF and D0 analysis as reported in
Ref.~\cite{Tevatron}.

\subsection{Methodology}

Experimental collaborations have reported various observed signal strengths in
the $i^{th}$ channel in terms of $\hat{\mu}_i = \sigma_i^{obs}/
\sigma_i^{SM}$, with $\sigma_i$ as the respective uncertainty. Here,
$\sigma_i^{obs}$ refers to the observed signal cross-section for a particular
Higgs mass, while $\sigma_i^{SM}$ is the signal cross-section for an SM Higgs
with the same mass. We calculate the corresponding values of $\mu_i$ for
various points in the space spanned by the parameters in terms of which we
have tried to capture the departure of the Higgs interactions from their SM
values, as explained in the previous section. One can express $\mu_i$ as

\be
\mu_i = {R^{prod}_i}\times R^{decay}_i /  R^{width}
\ee

\noindent where $R^{prod}_i, R^{decay}_i$ and $R^{width}$ are the factors
modifying the corresponding SM production cross-sections, decay width in a
particular channel, and the total decay width of the Higgs respectively.

The relevant production mechanisms at the LHC and Tevatron for the various
channels are listed below:

\begin{itemize}
\item For $\gamma\gamma$(inclusive), $ZZ^* \rightarrow 4\ell, WW^* \rightarrow
  \ell \ell \nu \nu$ and $\tau^+\tau^-$, one has to include all the production
  processes, namely, gluon fusion, vector boson fusion, associated $WH$ and
  $ZH$ production, and Higgsstrahlung from top-antitop pairs.

\item Only the associated $WH$ and $ZH$ production channels can lead to
  $b\bar{b}$ final states that can be separated from backgrounds.

\item VBF and GF are the dominant production modes for the $\gamma\gamma jj$
  channel reported by CMS; in particular, the contribution of VBF dominates
  once the appropriate cuts to identify forward tagged jets are imposed. This
  is reflected in the corresponding efficiencies: 15\% for VBF, and 0.5\% for
  GF~\cite{CMS-ggjj}.
\end{itemize}

In the most general case, the production cross-sections in the gluon fusion
(GF), associated production with a Z (ZH), associated production with $W^\pm$
(WH) and associated production with $t \bar{t}$ ($t \bar{t} H$) are modified
by the factors $x_g^2 \alpha_u^2$ ($R_{GF}$), $\beta_Z^2$ ($R_{ZH}$),
$\beta_W^2$ ($R_{WH}$) and $\alpha_u^2$ ($R_{t \bar{t}H}$) respectively.  In
the vector boson fusion (VBF) channel, the corresponding factor is given by

\begin{equation}
R_{VBF} \simeq \frac{3 \beta_W^2 + \beta_Z^2}{4} 
\end{equation}

The factor $R_{VBF}$ requires some explanation. In order to obtain this, first
of all, we have used the fact that the interference of the WW-fusion and the
ZZ-fusion diagrams is of the order of $1 \%$, and can therefore be ignored in
our calculation. Secondly, the WW-fusion contribution to the total
cross-section is roughly 3 times that of the ZZ-fusion contribution. For
details on this point we refer the reader to Ref.~\cite{Anatomy}. We have
cross-checked these facts for the LHC energies using the VBF@NNLO code of
Bolzoni {\it et al}~\cite{VBF-NLO}.

Note that these factors can only be used for inclusive channels of Higgs
search, whereby the efficiencies for event selection are the same in various
production modes. For channels in which special kinematic selection criteria
are used, one has to also include the corresponding efficiency factors, as
explained below for the $\gamma \gamma j j$ channel included by CMS.

For the $\gamma \gamma j j$ channel reported by CMS, we have considered $VBF$
and $GF$ as the two most dominant production modes. As
reported by CMS, the overall acceptance times selection efficiency of the
dijet tag for Higgs boson events is $15 \%$ for VBF ($\xi_{VBF}$) and $0.5 \%$
for GF ($\xi_{GF}$). Therefore, for the VBF channel, the factor modifying the
SM production cross-section is given by

\begin{equation}
R_{Hjj} = \frac{\xi_{VBF} R_{VBF} \sigma^{SM}_{VBF}+\xi_{GF} R_{GF} \sigma^{SM}_{GF}}{\xi_{VBF}  \sigma^{SM}_{VBF}+\xi_{GF}  \sigma^{SM}_{GF}}
\end{equation}

Next, we consider the Higgs decay widths in the channels under study. The
Higgs decay widths in the $ZZ^{*}, WW^{*}, \tau \bar{\tau}, b \bar{b}, c
\bar{c}$ and $gg$ channels get multiplied by $\beta_Z^2, \beta_W^2,
\alpha_d^2, \alpha_d^2, \alpha_u^2$ and $x_g^2 \alpha_u^2$ respectively. In
the loop-induced $\gamma \gamma$ channel, the contribution due to the top
quark and W-boson loops (and a small contribution due to the bottom and tau
loops) to the Higgs decay width is modified by the factor given by

\begin{equation}
R_{\gamma \gamma} = x_{\gamma}^2 \frac{|\frac{4}{3}\alpha_u e^{i \delta} A^H_{1/2} (\tau_t)+\frac{1}{3}\alpha_d  A^H_{1/2} (\tau_b)+\alpha_d  A^H_{1/2} (\tau_\tau)+\beta_W A^H_1 (\tau_W)|^2}{|\frac{4}{3} A^H_{1/2} (\tau_t)+\frac{1}{3}  A^H_{1/2} (\tau_b)+  A^H_{1/2} (\tau_\tau)+ A^H_1 (\tau_W)|^2}
\end{equation}

where the relevant loop functions are given by~\cite{Anatomy}

\begin{align}
A^H_{1/2} (\tau_i) &= 2[\tau_i + (\tau_i -1) f(\tau_i)] \tau_i^{-2} \nonumber \\
A^H_1 (\tau_i) &= -[2 \tau_i^2 + 3 \tau_i + 3(2 \tau_i-1)f(\tau_i)] \tau_i^{-2}
\end{align}

Here, $f(\tau_i)$, for $\tau_i \leq 1$ is expressed as,

\begin{equation}
f (\tau_i) = (\sin^{-1} \sqrt{\tau_i})^2 
\end{equation}

while, for $\tau_i > 1$, it is given by

\begin{equation}
-\frac{1}{4} \left[ \log \frac{1+\sqrt{1-\tau_i^{-1}}}{1-\sqrt{1-\tau_i^{-1}}} -i \pi \right]^2
\end{equation}

In the above equations $\tau_i$ denotes the ratio $m_H^2 / 4 m_i^2$. 

With the values of $\mu_i$ thus calculated for various points in the parameter
space, we first obtain the best fit values for these parameters (upto 7 at a
time) corresponding to the global minimum of the function $\chi^2$, defined as

\be
\chi^2 = \sum_i \frac{(\mu_i - \hat{\mu_i})^2}{\sigma_i^2}
\ee  

Note that, the experimental collaborations, in some cases, have reported
asymmetric error bars on the data. In order to include such error bars in the
above definition, we use the following prescription. If $(\mu_i - \hat{\mu_i})
> 0$, we use the positive error bar $\sigma_i^{+}$, while if $(\mu_i -
\hat{\mu_i}) < 0$, we use the negative error bar
$\sigma_i^{-}$~\cite{James}. Note that, in cases for which we have combined
more than one experimental data points to obtain a single input data, we
obtained the average signal strength $\bar{\hat{\mu}}$ and the corresponding
uncertainty $\bar{\sigma}$ using the following relations:

\begin{align}
\frac{1}{\bar{\sigma}^2} &= \sum_i \frac{1}{\sigma_i^2} \nonumber \\
\frac{\bar{\hat{\mu}}}{\bar{\sigma}^2} &= \sum_i \frac{\hat{\mu_i}}{\sigma_i^2}
\end{align} 

This method has been used, for example, in case of CMS data on inclusive
$\gamma \gamma$ and $\gamma \gamma j j$ channels, where the results were
presented in four different categories in the former case while with both
loose and tight dijet tags in the latter case~\cite{CMS-ggjj}.

We have also used the above method in combining different contributions to the
theoretically calculated $\mu$ values, for example, while combining the
contributions of $WH \rightarrow l \nu b \bar{b}$, $ZH \rightarrow l^+l^- b
\bar{b}$ and $ZH \rightarrow\nu \bar{\nu} b \bar{b}$ to associated Higgs
production with gauge bosons and the subsequent decay of the Higgs to a bottom
pair~\cite{LHCHWG,Baglio:2010um}. Such combinations were necessitated by the fact 
that the experimental collaborations have reported a single signal strength value 
in the $b \bar{b}$ channel.

After obtaining the best-fit values for the parameters by minimizing the
$\chi^2$ function, we consider two-dimensional contours for various pairs of
them about the global minimum, fixing the remaining ones at their best-fit
values. The contours are drawn for 68.3\% and 95.4\% confidence intervals. One
important point in which our study differs from most earlier ones is that, in
the cases where these two-dimensional contours are drawn, {\em the remaining
  parameters are fixed not at their SM values but at values corresponding to
  the global minimum of the $\chi^2$ fit.} The various values of $\Delta
\chi^2$ for the 68.3\% and 95.4\% confidence intervals are tabulated, for
example, in Refs.~\cite{minuit, NR}.

\section{Results}

We obtain the global fits in three different cases. In the first case (case
A), the phase $\delta$ is set to zero, and the seven remaining parameters are
varied across the pre-decided ranges, as described in section 2. In the second
case (case B), we try to see the effects of the phase, and set $\beta_W =
\beta_Z$ in the process. The same exercise is done in the third case, but with
the phase set to zero again.  The simplification in the last two cases is done
because $\beta_W \ne \beta_Z$ amounts to the breaking of custodial SU(2) at
the tree-level, and is subject to rather stringent constraints from precision
electroweak observables.  Though we go beyond these constraints in the first
case, assuming the simultaneous existence of more than one gauge-invariant
effective interactions at the same time, we desist from such speculative
exercise in the cases B and C, and keep $\beta_W = \beta_Z$ within the most
stringent precision electroweak limits. It should be noted that the phase
$\delta$ does not affect the loop-induced contribution to the T-parameter,
since it cancels in the relevant self-energy diagrams.  Table 2 contains the
best fit values of various parameters in the three cases.

\begin{table}[t]
\centering
\begin{tabular}{|l|c|c|c|c|c|c|c|c|}
\hline
Case & $\beta_W$ & $\beta_Z$ & $\alpha_u$ & $\alpha_d$ & $x_g$ & $x_{\gamma}$ & $\epsilon$ & $\delta$ \\

\hline
A & 1.2 & 1.6 & -0.6 & -1.2 & 1.8 & 1.3 & 0.6 & $0^{*}$ \\

\hline

B & 1.06 & 1.06 & -1.05 & 0.95 & 0.8 & 1.0 & 0.2 & 0.55 \\ 

\hline

C & 1.07   &1.07   &-1.3   &-0.96   &0.65   &0.95  &0.2  & $0^{*}$\\
\hline
\end{tabular}
\caption{Best-fit values of the various parameters in the three cases
  considered. In cases A and C, $\delta$ has been fixed at $0$ (indicated with a
  '$*$'). In cases B and C, the relation $\beta_W = \beta_Z$ has been imposed, and
  their values have been restricted within precision constraints.}
\label{tab:tab2}
\end{table}

\begin{figure}[h!]
\begin{center}
\vspace*{-0.5cm}
\centerline{\epsfig{file=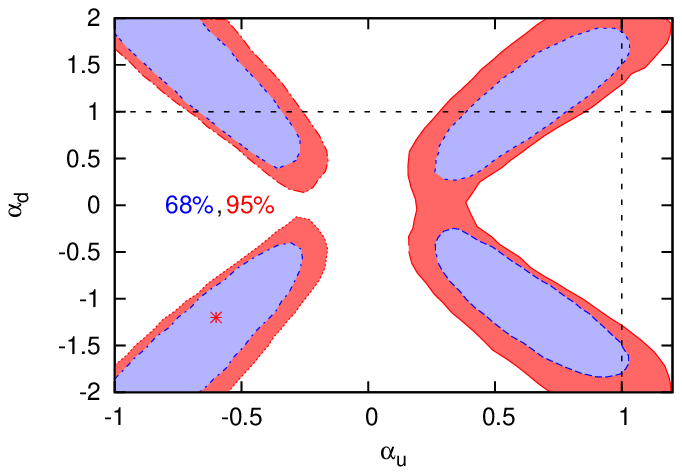,width=7.5cm,height=5.7cm,angle=-0}
\hskip 20pt \epsfig{file=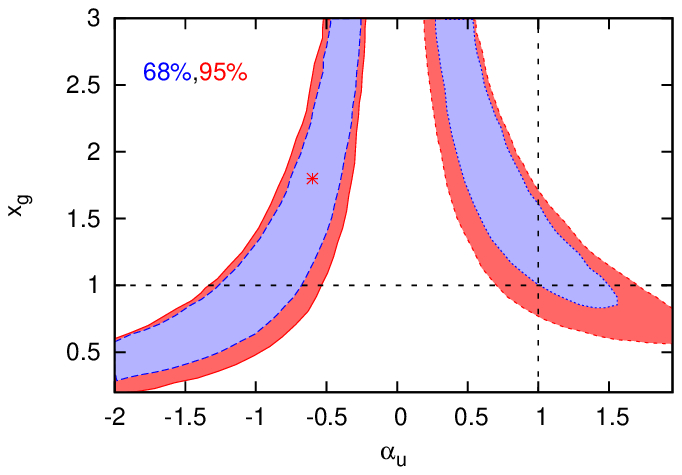,width=7.5cm,height=5.7cm,angle=-0}}
\centerline{\epsfig{file=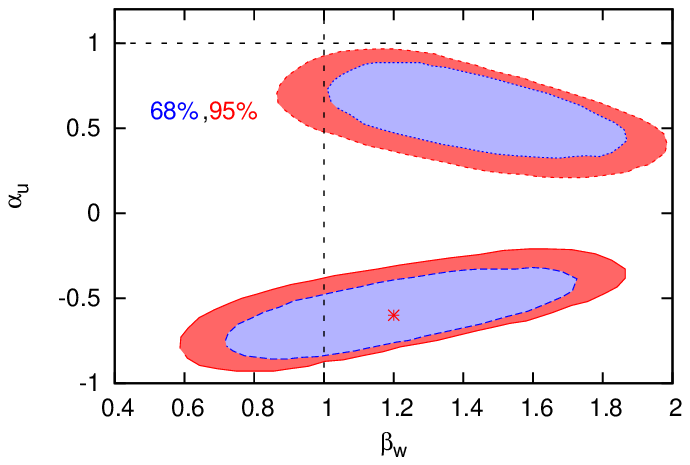,width=7.5cm,height=5.7cm,angle=-0}
\hskip 20pt \epsfig{file=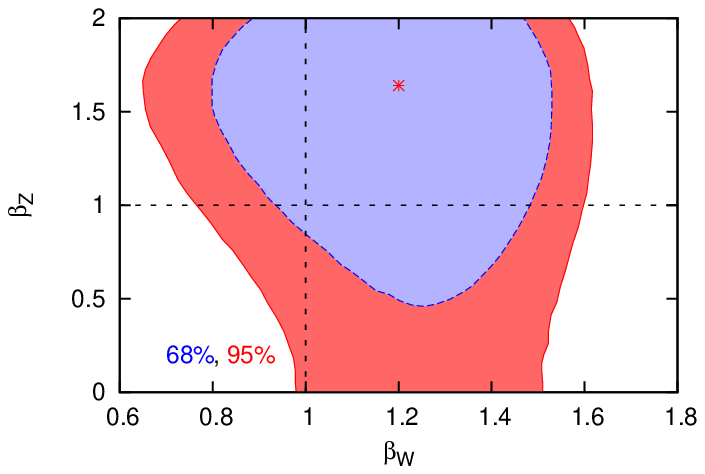,width=7.5cm,height=5.7cm,angle=-0}}
\centerline{\epsfig{file=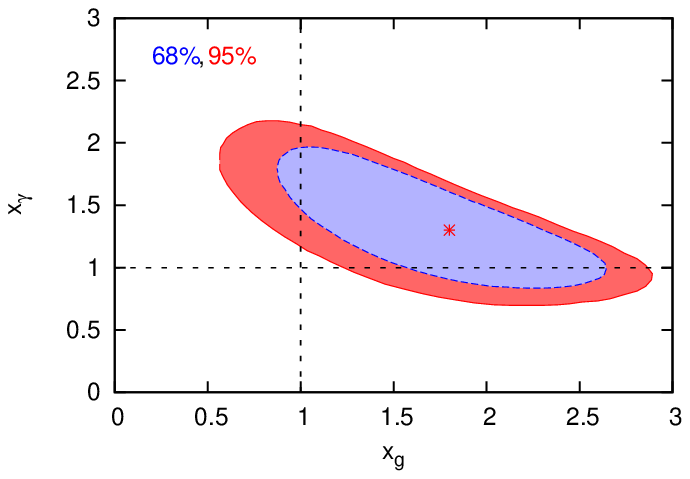,width=7.5cm,height=5.7cm,angle=-0}}
\caption{Two-dimensional contour plots for $68\%$ and $95\%$ confidence
  intervals, for case A, with rest of the parameters fixed at their best-fit values. The
  best-fit point is also marked separately by a '$*$'. In this case $\delta$ has been
  fixed at $0$, whereas $0\leq \beta_W,\beta_Z\leq2.0$, and $\beta_W \neq
  \beta_Z$.}
\label{fig:no-delta}
\end{center}
\end{figure}
\begin{figure}[h!]
\begin{center}
\vspace*{-0.5cm}
\centerline{\epsfig{file=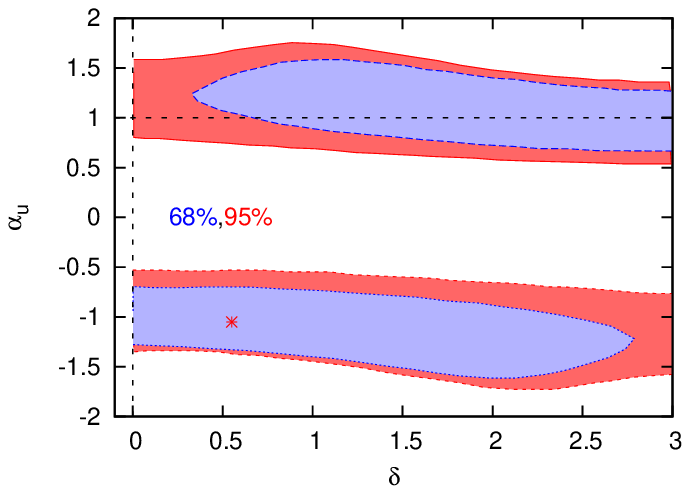,width=7.5cm,height=5.7cm,angle=-0}
\hskip 20pt \epsfig{file=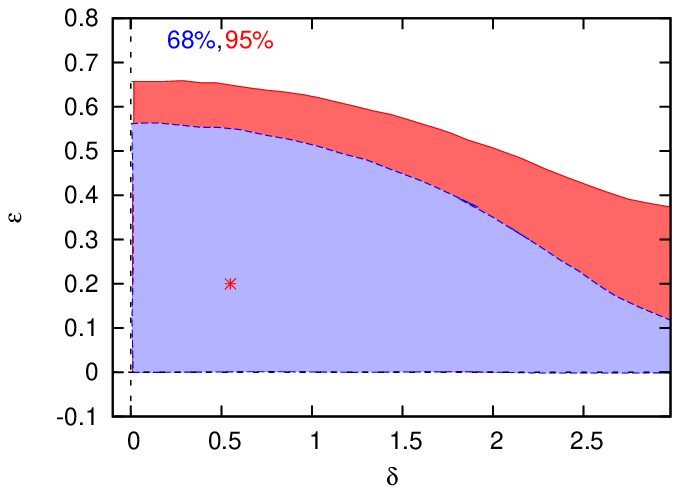,width=7.5cm,height=5.7cm,angle=-0}}
\centerline{\epsfig{file=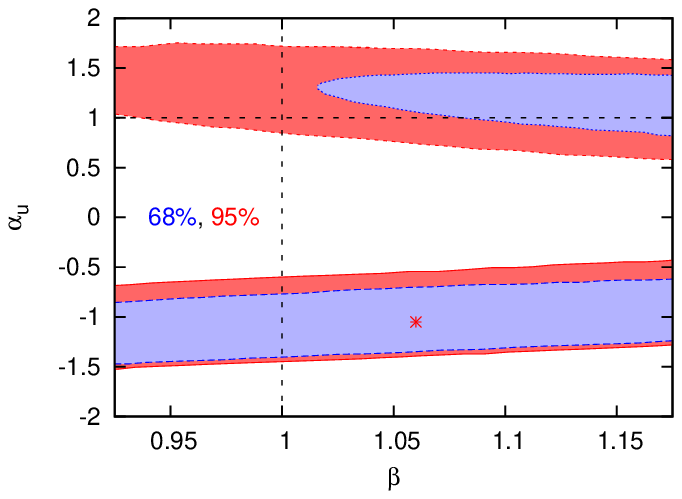,width=7.5cm,height=5.7cm,angle=-0}
\hskip 20pt \epsfig{file=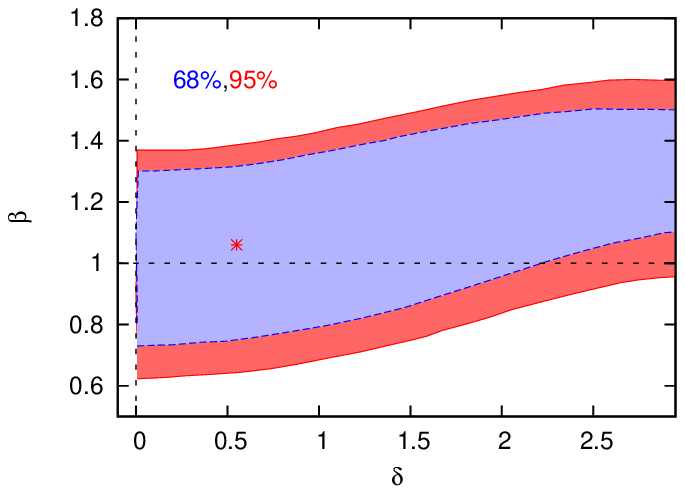,width=7.5cm,height=5.7cm,angle=-0}}
\centerline{\epsfig{file=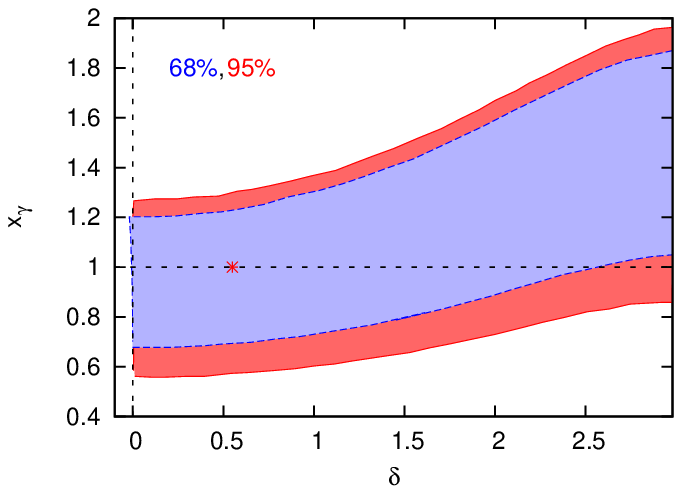,width=7.5cm,height=5.7cm,angle=-0}
\hskip 20pt \epsfig{file=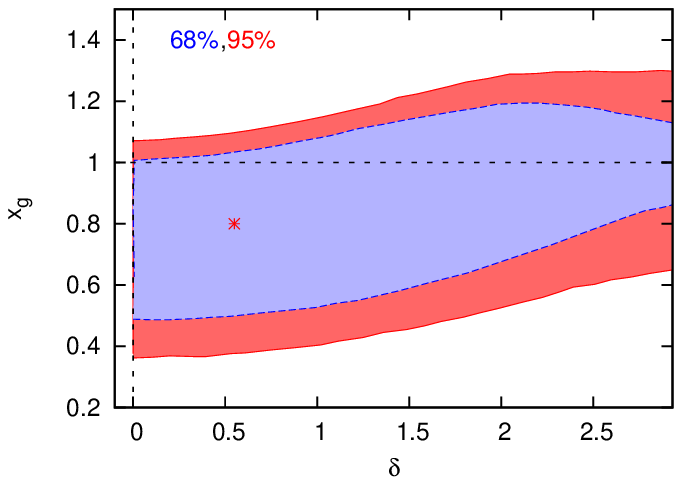,width=7.5cm,height=5.7cm,angle=-0}}
\caption{Two-dimensional contour plots for $68\%$ and $95\%$ confidence
  intervals, for case B, with rest of the parameters fixed at their best-fit
  values. The best-fit point is also marked separately by a '$*$'. In this
  case $\delta$ has been varied in the range $\{0,\pi\}$, whereas $0.92\leq
  \beta\leq 1.18$, with $\beta \equiv \beta_W =\beta_Z$.}
\label{fig:delta-1}
\end{center}
\end{figure}

The obtained best-fit values for the parameters have been tabulated in
Table~\ref{tab:tab2}. The minimum values of $\chi^2$, $\chi^2_{min}$ are
$6.19$, $7.03$ and $7.03$ for cases A, B and C respectively. Note that, the
number of degrees of freedom (for cases A and B) in our global fits is $7$, 
since we have $14$ input data points, and $7$ parameters. However, for case C
the number of degrees of freedom is $8$, since $\beta_W = \beta_Z$ and $\delta=0$.
We present the two-dimensional 68.3\% and 95.4\% confidence interval contours for various
pairs of parameters about the global minimum, for cases A and B, in 
Figures~\ref{fig:no-delta} and \ref{fig:delta-1} respectively. As mentioned 
before, while drawing these contours, we have fixed the remaining parameters 
at their best-fit values. In this connection, we note that the consistency 
of our methodology has been checked by also performing a two parameter fit 
of the data (with the parameters pertaining to the overall modifications of 
the fermion and gauge couplings), and our results in that case agree with 
those presented in recent literature~\cite{fits-recent}.

Let us first consider the case with $\delta = 0$ and $\beta_W \ne \beta_Z$. 
First of all, there is a global tendency in the experimental results reported 
so far towards best fit values of the WW rates that are consistent with a 
shortfall compared to the SM prediction. For ZZ, on the other hand, there is 
an excess for ATLAS and a shortfall, but less than that for WW, for CMS. Though 
these are all consistent at the 1$\sigma$ level, the least-square minimisation 
inevitably yields a best fit for $\beta_Z$ higher than that of $\beta_W$.  
Since this causes one to go beyond the most stringent of precision constraints, 
one has to admit that such a best fit central value favours the simultaneous 
presence of more than one gauge-invariant anomalous WW/ZZ interactions for the 
Higgs. It should, of course, be kept in mind that one has to wait for the 
accumulation of more data before a final verdict can be spelt on this. In 
cases B and C, where we set $0.92\leq \beta \leq 1.18$, with 
$\beta \equiv \beta_W = \beta_Z $, we find the best-fit values of $\beta$ to 
be close to $1$.

As far as the best fit values for the parameters multiplying the fermion
couplings are concerned, we find that $\alpha_u$ receives a sign relative to
$\beta_W$, with a value $-0.6, -1.05$ and $-1.3$ for cases A, B and C
respectively. This relative sign between the top-quark Yukawa coupling and the
W-boson coupling to the Higgs can be traced to the fact that the observed
excess in the $\gamma \gamma$ mode by both the ATLAS and CMS experiments can
be explained by a constructive interference between the loop amplitudes due to
the top and the W-loop. In contrast to $\alpha_u$, the magnitude of the
best-fit value of $\alpha_d$ in all cases turns out to be close to unity. Its
sign does not affect the Higgs production cross-sections and decay branching
ratios much, except for a small effect in the $\gamma \gamma$ decay
mode. However, to be completely general, we varied it in the range $-2.0$ to
$2.0$. One should note that, as can be seen in Figure~\ref{fig:no-delta}, the
$68 \%$ and $95 \%$ contours allow for both positive and negative values of
$\alpha_d$, and these contours bring out the global picture more clearly than
just the best-fit values. It should be noted that $\alpha_d$ has a positive
best fit value only when the phase $\delta$ is turned on. This elicits an
interesting feature of the global fit, when the phase in fermion couplings is
included. We also note that, from contours involving $\alpha_u$ and $\alpha_d$
in Figure~\ref{fig:no-delta} and $\alpha_u$ in Figure~\ref{fig:delta-1}, it is
clear that values close to zero of these two parameters are clearly
disfavoured. This, therefore, should severely constrain fermiophobic models
for the Higgs~\cite{fphobia}.

We also find the best-fit values for $x_g$ and $x_{\gamma}$ to differ, and at
the same time lie away from their SM value of $1.0$. This suggests that not
only is the presence of new coloured or electrically charged states coupling
to the Higgs plausible, but that the number of states contributing to the
gluon fusion process can be different from those contributing to the
$\gamma\gamma$ decay mode.

\begin{figure}[h!]
\begin{center}
\vspace*{-0.5cm}
\centerline{\epsfig{file=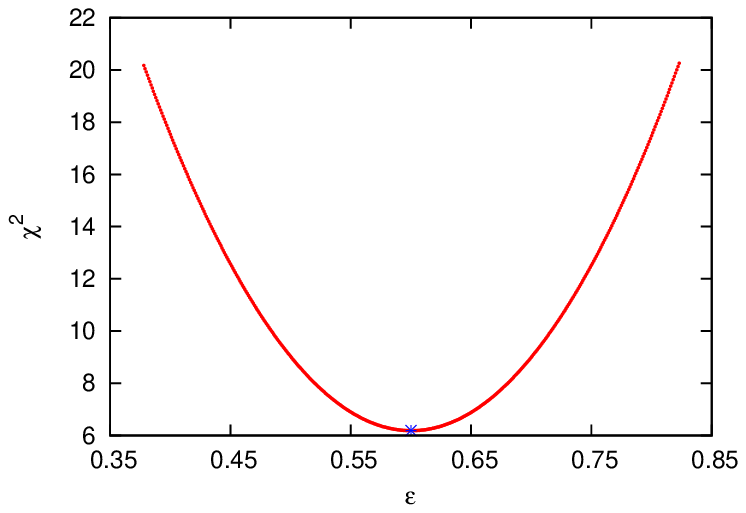,width=7.5cm,height=5.7cm,angle=-0}
\hskip 20pt \epsfig{file=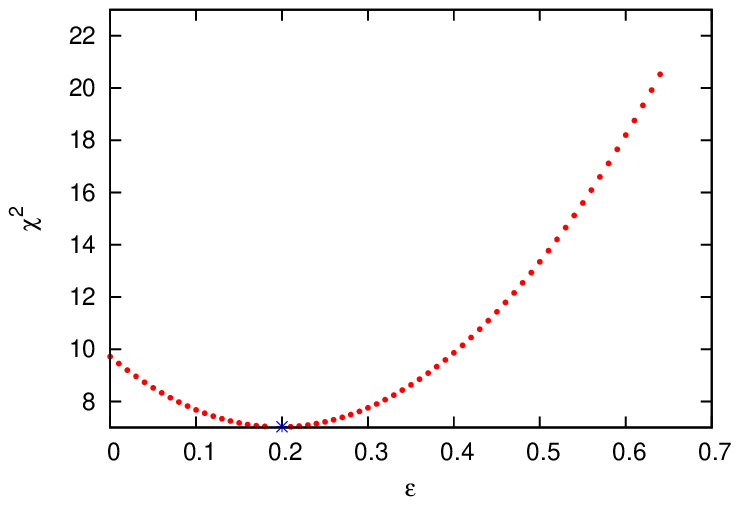,width=7.5cm,height=5.7cm,angle=-0}}
\caption{Variation of the $\chi^2$ function with the invisible branching
  fraction of $H$ ($\epsilon$) in cases A (left panel) and B (right panel). In
  case A, $\delta=0$ and $\beta_W \ne \beta_Z$, whereas in case B, $\delta$
  has been varied in the range $\{0,\pi\}$ and $0.92\leq \beta\leq 1.18$, with
  $\beta \equiv \beta_W =\beta_Z$. }
\label{fig:epsilon}
\end{center}
\end{figure}

In case A, we find a rather large allowed value for the invisible branching
ratio of the Higgs boson. This may be due to the fact that the production rate
in the gluon fusion channel, as well as the $Z$ boson contribution to the
associated production and VBF processes (as the best-fit of $\beta_Z$ is
$1.6$) are enhanced. This creates room for the simultaneous enhancement of
certain visible modes ($\gamma \gamma, ZZ^*$) as well as the existence of
substantial invisible branching fraction. Note that we observe a reduction in
$\epsilon$ once $\beta_W$ and $\beta_Z$ are set to be equal, and in the range
allowed by electroweak precision data (cases B and C), the best-fit value in
both these cases being $0.2$. Thus our conclusion is that, while invisible
decays of the Higgs are more constrained once a correlation between the $HWW$
and $HZZ$ couplings is enforced, the overall freedom in a seven-parameter
space nonetheless keeps enough scope for such decays. We demonstrate the
variation of the $\chi^2$ function with $\epsilon$, in cases A and B, with
rest of the parameters fixed at their best-fit values, in
Figure~\ref{fig:epsilon}.

\begin{figure}[h!]
\begin{center}
\vspace*{-0.5cm}
\centerline{\epsfig{file=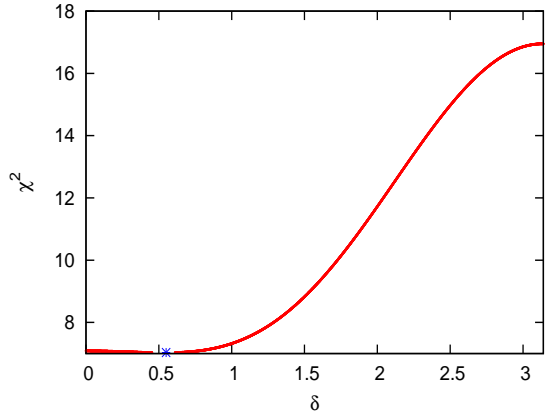,width=7.5cm,height=5.7cm,angle=-0}}
\caption{Variation of the $\chi^2$ function with the phase in the up-type quark Yukawa
  coupling, $\delta$, in case B. In this case $\delta$ has been varied in the range
  $\{0,\pi\}$, whereas $0.92\leq \beta\leq 1.18$, with $\beta \equiv \beta_W =\beta_Z$.}
\label{fig:delta-2}
\end{center}
\end{figure}

The interesting feature to be noted is that the phase $\delta$ is allowed to
assume values as large as 0.55 in the best fit. Once this non-trivial phase is
there, the 95\% confidence level contours of all other anomalous interactions
tend to include the SM values. It should however be also noted from 
Figure~\ref{fig:delta-2} that although the best fit value for $\delta$ is 0.55, the $\chi^2$
function is very slowly varying in the entire range between 0 and 0.6.  The
curve as a whole indicates that the best fit value is relatively unspecified
between 0 and 1, and we have to wait for more data to have a clearer picture
about the occurrence of this phase. Also, a comparison between the $\alpha_u-\beta_W$ 
plot in Figure~\ref{fig:no-delta} and the $\alpha_u-\beta$ plot in Figure~\ref{fig:delta-1} 
shows that the contours not only shift but have also bigger coverage in the $\delta \ne 0$ case.

\section {Summary and conclusions}

The attempt in this study is to see how much scope of new physics is contained
within the data on Higgs search, taking into account not only the results
available from 7 and 8 TeV runs of the LHC but also the full data set of the
Tevatron. We have taken a completely model-independent stand, without any bias
of correlation between the anomalous Higgs couplings of $T_3 = + 1/2$ and
$-1/2$ fermions, as well as the W-and Z-couplings. We have in addition
accounted for additional states contributing to the effective interaction of
the Higgs with photon and gluon pairs. Since such contributions can come
through different sets of states for the $gg$ and $\gamma\gamma$ pairs, we
assign a separate uncorrelated parameter for each of these
processes. Furthermore, we have taken into account an arbitrary phase in the
top-pair coupling with respect to that from a W-pair, which can in principle
non-trivially affect the loop-induced decay $H\longrightarrow
\gamma\gamma$. Last but not the least, a non-vanishing width for the Higgs
decaying into invisible final states is kept as a free parameter in our global
fit.

Our study takes into account not only the contributions to all final states
from the dominant gluon fusion channel but also the VBF and all associated
production subprocesses, with appropriate weightage to the efficiencies for
different channels. Also, asymmetric uncertainties in the data, as reported by
the experiments in some cases, have been used in our analysis. The present study is
nonetheless based on certain simplifying assumptions. These include, for example, 
the same cut efficiencies taken in inclusive final states where more than one 
production channels are involved, which can be improved as and when the efficiencies
are published by the experimental collaborations.

The global fits for the minimum of $\chi^2$ over seven-parameter scans, for
three sets of combinations, yield the best fit values of each. Subsequently,
68\% and 95\% confidence level contours for various parameters have been
presented, where all other parameters have been kept at their global best fit
values (in contrast to most studies where they are fixed at the corresponding
SM values).

The most important conclusion we draw is that it is too early to say
that signals reveal {\em the standard model Higgs}, since the $2\sigma$ 
contours allow departure from the SM values. A few general trends that
show up are

\begin{itemize}
\item A fermiophobic Higgs is by and large disfavoured.
\item There is in general the hint in the best-fit values of a relative sign between
the couplings to the up-type fermion and the gauge boson pairs.
\item A non-trivial phase in the top-quark coupling can have 
a rather important role. Interestingly, the SM values of the
remaining parameters tend to get included within the $2\sigma$
contours once this phase is turned on. 
\item It still seems possible to accommodate an invisible decay width of the
  Higgs. In particular, it can be as large as 60\% if one allows a breakdown
  of the custodial SU(2) symmetry.  Though this is {\em prima facie}
  restricted by electroweak precision data, it remains to be checked whether
  the values of $\beta_W \ne \beta_Z$ can be made consistent with precision
  data, by introducing more than one gauge invariant higher-dimensional
  operators at the same time.  However, invisible branching ratios upto 20\%
  or more are possible even without such manoeuvres.
\end{itemize}

Data from the LHC are currently in a state of flux, and therefore
the  numerical results on which our analysis is based can change
with growing statistics. However, while the trends pointed out by us 
might change as more accurate data become available, the rather general
approach used by us should continue to serve as a template for future analyses.

\section*{Acknowledgement}
We thank Satyaki Bhattacharya, Frederick James and Lorenzo Moneta for many
valuable discussions and useful correspondence. We would also like to thank
Dhiraj Kumar Hazra for an important computational help. S.B. and B.M. thank
the Indian Association for the Cultivation of Science, Kolkata, for
hospitality while this study was in progress. This work was partially
supported by funding available from the Department of Atomic Energy,
Government of India for the Regional Centre for Accelerator-based Particle
Physics, Harish-Chandra Research Institute. Computational work for this study
was partially carried out at the cluster computing facility in the
Harish-Chandra Research Institute (http:/$\!$/cluster.hri.res.in).


\end{document}